# Uncertainty of Supply Chains: Risk and Ambiguity[1]

## d'Artis Kancs

Directorate-General Joint Research Centre, European Commission;
Science Research Innovation Implementation Centre, Latvian National Armed Forces

d'artis.kancs@ec.europa.eu

## *ABSTRACT*

Motivated by the recently experienced systemic shocks (the COVID-19 pandemic and the full-fledged Russia's war of aggression against Ukraine) – that have created new forms of uncertainties to our supplies – this paper explores the supply chain robustness under risk aversion and ambiguity aversion. We aim to understand the potential consequences of deeply uncertain systemic events on the supply chain resilience and how does the information precision affect individual agents' choices and the chain-level preparedness to aggregate shocks. Augmenting a parsimonious supply chain model with uncertainty, we analyse the relationship between the upstream sourcing decisions and the supply chain survival probability. Both risk-averse and ambiguity-averse individually-optimising agents' upstream sourcing paths are efficient but can become vulnerable to aggregate shocks. In contrast, a chain-level coordination of downstream firm sourcing decisions can qualitatively improve the robustness of the entire supply chain compared to the individual decision-making baseline. Such a robust decision making ensures that in the presence of an aggregate shock – independently of its realisation – part of upstream suppliers will survive and the final goods' supply will be ensured even under the most demanding circumstances. Our results also indicate that an input source diversification extracts a cost in foregone efficiency.

---

[1] The authors acknowledge helpful comments from Etienne Vincent as well as participants of the NATO Computer Assisted Analysis, Exercise, Experimentation (CA2X2) Forum in Rome organised by the NATO Modelling and Simulation Centre of Excellence; ZPIIC seminar in Pinki, NATO Innovation Challenge finalists presentations in Bucharest organised by the NATO Innovation Hub the NATO Supreme Allied Commander Transformation; NATO Operations Research and Analysis (OR&A) Conference in Copenhagen organised by the NATO Supreme Allied Commander Transformation and NATO Science and Technology Organization; NATO Emerging and Disruptive Modelling and Simulation Technologies Symposium in Bath organised by Modelling and Simulation Group, NATO Science and Technology Organization. The authors are solely responsible for the content of the paper. The views expressed are purely those of the authors and may not in any circumstances be regarded as stating an official position of the NATO, European Commission or the Latvian National Armed Forces. Any remaining errors are solely ours.

## 1.0 INTRODUCTION

Firms rely on upstream supply chains to source the intermediate inputs that are needed for production. As experienced recently, global supply chain (GSC) can be disrupted by natural disasters, pandemics, wars, changes in policy regimes, or transportation raptures. Not only the fragility of global supply chains is raising to historically high levels since several years, but also the landscape of hybrid threats is expanding, including new forms such as the weaponisation of supply chains (RUSI 2023).[2] For policy makers, the supply chain resilience and robustness is important for at least two reasons. First, singular upstream shocks can propagate to the rest of the economy through input-output linkages, resulting in aggregate macroeconomic fluctuations. Second, ensuring resilient supplies in critical sectors, such as food, water and energy, is of a strategic importance for the entire society. The currently less and less predictable global security environment is requiring the Alliance to increase the focus on preparedness and resilience.[3] Building a system-wide resilience should ensure that the socio-politico-economic fabric can function in the face of adversity under a wide range of contingencies while ensuring the critical supplies.[4] To understand the chain-level implications of resilience-enhancing and shock-mitigation actions at the individual and societal-level, we study the mechanics of firm-level upstream sourcing decisions on the aggregate preparedness and chain-level robustness.

Two supply chain developments with a global character and dynamically interrelated across industries and countries have accelerated in recent years. One is an increasing fragmentation and interdependency of global production networks both across upstream tiers as well across different supply chains (Antras and Chor 2022). In the age of globalisation of complex cross-border production networks, lower cost is usually associated with economies of scale in input sourcing, whereas higher quality inputs tend to be found in markets with a niche expertise – both driving company participation in GSCs (Baldwin and Freeman 2022). On the downside, the specialisation and cost advantages that arise from involvement in GSCs are unavoidably associated with greater uncertainty in the face of shocks, such as global pandemics, the climate crisis and adversarial attacks (Todo et al. 2023). In the same time, the landscape of hybrid threats is expanding including new forms such as the weaponisation of global supply chains (Hybrid CoE 2023).[5] Recent examples include the Russia's weaponisation of energy supply to Europe and "blackmailing" the world by holding food supplies hostage as part of the war strategy. Using a hybrid tactics, the authoritarian regimes and strategic challengers test the Alliance's resilience and seek to exploit the openness, interconnectedness and digitalisation, interfere in democratic processes and institutions, and target the security of citizens through a hybrid tactics (Hybrid CoE 2023).

Both global realities – the fragmentation of global input sourcing and new domains of hybrid warfare – reveal significant vulnerabilities of our supply chains which can magnify and accelerate impacts of system-wide threats to the Alliance. As acknowledged by the Secretary General Stoltenberg: "over-reliance on the import of key commodities, like energy [on the sourcing-side, and] exporting advanced technologies, like Artificial Intelligence [on the selling-side] can create vulnerabilities and weakened resilience". The challenge the decision makers face is to achieve long-term security goals, ensure resilient and diversified supply chains in place while allowing for a continued flow of essential goods and avoid shortages in the short-, medium- and long-run without neglecting the economic needs of business and society. Our study contributes to better understanding of interdependencies, mechanics and trade-offs the socio-politico-economic fabric may be facing during a transition to robust and resilient critical supply chains.

---

[2] Royal United Services Institute (2023) www.rusi.org/explore-our-research/publications/commentary/uk-needs-confront-weaponisation-nuclear-supply-chain

[3] For the purpose of the present study, the Alliance is defined as the 31 NATO's member countries plus Individually Tailored Partnership Programme (ITPP) countries (currently, Australia, Japan, New Zealand and South Korea).

[4] www.nato.int/docu/review/articles/2019/02/27/resilience-the-first-line-of-defence/

[5] European Centre of Excellence for Countering Hybrid Threats (2023) www.hybridcoe.fi/publications/hybrid-threats-a-comprehensive-resilience-ecosystem/

We investigate the how-to-enhance-resilience question formally by augmenting a parsimonious supply chain model with uncertainty and simulating alternative upstream sourcing strategies in the presence of information incompleteness. Specifically, we study the relationship between the individual upstream supplier concentration (which in upper tiers may not visible to the downstream manufacturer) and the aggregate supply chains' robustness (survival probability). In a number of counterfactual scenarios we simulate alternative sourcing strategies in the presence of uncertainty – the shock realisation is unknown to firms at the time of making sourcing and production decisions. The model depicts shocks in a reduced form as simply regimes in which the labour force is subject to exogenous ruptures (e.g. the non-pharmaceutical interventions and the health effects of the COVID-19 pandemic on both the raw labour and human capital). Frist, we investigate the supply chains' robustness outcomes when firms optimise upstream sourcing decisions individually in market economies. We show that a cost optimisation and profit maximisation can drive all firms to source from one least cost location with most productive upstream suppliers. If hit by an aggregate shock, the entire labour supply in this location would abruptly decrease to zero, implying a disruption of the entire supply chain (as at least one firm is required at every tier to produce the final good). In a second set of simulations we investigate the role a central planner- (government) can play to mitigate the supply chain's vulnerability to aggregate shocks and enhance resilience. We show that by re-incentivising the upstream supplier diversification decisions in the presence of chain-level externalities, the supply chain's survival probability can be maximised. While the central planner's first-best solution features a robust sourcing path, individual firms do not internalise the benefits of the chain-level robustness and under-diversify upstream suppliers in laissez-faire market economies.

Our analysis contributes both to the resilience literature focusing on critical and strategic supplies as well as to a broader efficiency-resilience debate in the context of policy making under deep uncertainty. As our first contribution, we apply the multiple-priors framework to supply chains facing not only risk, but also ambiguity. Ambiguity averse agents are not confident enough to assign probabilities to shock events about which they have little information, such as the COVID-19 pandemic. Instead, uncertainty about probability assessments is described by the size of the agents' belief set, as in robust statistics (Maronna et al. 2018). When ambiguity averse agents evaluate probabilities of shock events, they act as if they are relatively pessimistic, and more so the larger their belief set. Such a strategy guarantees the survival of the supply chain. As our second contribution, we study how ambiguity-averse decision makers can practically enhance resilience and ensure a chain-level robustness. Our results imply that an input source diversification extracts a non-trivial cost in foregone efficiency. A supply chain path with greater resilience to shocks has to sacrifice efficiency during 'normal' times. Differently from the previous literature, our focus is on the question if, under what uncertainty environment and how much resources in terms of a foregone efficiency would a decision maker be willing to divert to an increased robustness.

For policy makers, our results suggest that achieving a socio-politico-economic resilience that meets the Alliance's baseline requirements – which must be maintained under the most demanding circumstances – will require a mobilisation of resources. For example, efforts to restructure supply chains by enhancing resilience and robustness may have to be accompanied by financial incentives to minimise the company costs associated with an upstream supplier diversification. As noted by Secretary General Stoltenberg at the World Economic Forum 2022: "we should not trade long-term security needs for short-term economic interests",[6] which implies costs and sacrifices. Our study contributes to increasing transparency both about the policy effectiveness in terms of enhancing resilience and efficiency in terms of societal costs in foregone efficiency. The present study builds on and complements the existing Alliance's Science & Technology Organisation (STO) analytical support, including the Multi-Dimensional Data Farming, Causal Reasoning, and resilience tools. The Resilience Data Analytics Tool is developed to assess the levels of resilience by leveraging open-source data, big data analytics, machine learning, and data visualisation and allows the identification of potential shocks. The Resilience Model provides a holistic framework for simulating a wide range of Political, Military, Economic, Social, Information, and Infrastructure (PMESII) shocks, such as electricity blackout, cyber attack, martial law enforcement, big human movement, state of war, armed conflict), and allows assessing several resilience domains.

---

6 www.nato.int/cps/en/natohq/opinions_195755.htm

## 2.0 GLOBAL SUPPLY CHAINS: EVIDENCE

We begin by taking stocks of the Alliance's foreign input dependence. Knowing economies' international exposure on intermediates input and output markets is crucial to identify potential vulnerabilities and allow for policy actions where needs arise before major GSC-disruptions occur. We approach this question by querying recent statistical data through a macro lens. Specifically, a value-added approach is employed to assess how the industrial production is allocated internationally and how each stage of production contributes to the final product (Johnson 2018). To measure international GSCs dependencies, national Input Output (IO) tables are connected across borders using bilateral trade data to construct World Input-Output Tables (WIOT). These data are then applied to measure trade in value added, as well as the length and location of producers in GSCs. By acknowledging the relevance of GSCs in the world intermediate goods trade, the ultimate goal is to construct a reliable tool for the counterfactual analysis.

### 2.1 Foreign exposure of Alliance

An understanding of the Alliance's foreign exposure via the GSC channel requires a detailed knowledge of where are goods made. To answer this question, employing a value-added approach we look at how production is allocated internationally and how each stage of production contributes to the final product. We use the information contained in WIOT to shed light on value-added trade flows across countries and compute the implied degree to which production processes have become globalised. A variety of metrics has been developed to assess the foreign exposure of a sector or economy as a whole (see e.g. Johnson 2018). For example, the content of value added in final goods, value added in gross exports, positioning in GSCs or the Foreign Input Reliance (FIR), which measures the sourcing-side exposure of a sector or the entire economy. We use the Inter-Country Input-Output (ICIO) data from the OECD to compute FIR for seven largest Alliance's economies and the strategic challenger China in 2019 (the most recent available data). The computed bilateral FIR corresponds to the share of foreign sources used as intermediate inputs into domestic production.

Table 1, panel (a) reports row nations' reliance on inputs from column nation for the manufacturing production. Cell shades are indexed to share sizes; darker shades indicate higher FIR (more import-dependent).[7] For example, 11.8 in the row for Canada (CAN) and the column for China (CHN) indicates that 11.8% of Canadian manufacturing production was made using inputs sourced directly and indirectly from China. The global dominance of China in intermediate input trade can also be seen by the CHN column colour, which is shaded primarily in dark. The result that the CHN column is relatively dark indicates that China is an important supplier of inputs to manufacturing industries of all analysed Alliance's economies. It is also worth noting the asymmetry between the USA manufacturing production's reliance on Chinese inputs, 9.9%, and China's manufacturing production's reliance on US inputs, 3.7%.

Next, we investigate how Foreign Input Reliance has changed during the last two decades by comparing FIR in 2019 with FIR in 2000. Table 1, panel (b) reports change in row nations' reliance on inputs from column nation for manufacturing production between 2000 and 2019. Darker-shaded cells indicate larger changes in FIR. For most countries the bilateral FIR matrix with respect to China was considerably larger in 2019 than it was in 2000. In panel (b), the figures in the China column are all positive and all significantly different from zero, indicating that the seven largest Alliance's economies' input dependence on China has increased. In contrast, the figures for the USA column are small, mostly, under 1ppt, and some input share changes are even negative (e.g. China). Most of the panel (b) entries for other countries are negative. Overall, the reliance of Alliance on Chinese inputs has increased substantially between 2000 and 2019, whereas the opposite is observed for Chinas reliance on inputs from the seven largest Alliance's economies (last row in Table 1, panel (b)).

---

[7] The matrix diagonal elements are suppressed, as we are interested in foreign inputs and foreign exposure. The diagonal elements would show a nation's input reliance on itself - both in terms of direct domestic sourcing and indirect sourcing through the re-import of previously exported inputs.

Industries and countries participating in GSCs are exposed also to sales-side shocks. Therefore, it is likewise important to understand domestic industries' foreign dependence on the output side. Conceptually similar to the FIR index - which measures countries' total reliance on foreign production on the sourcing side - the Foreign Market Reliance (FMR) index measures countries' reliance on foreign markets on the sales side. Table 1, panel (c) reports row nations' total input sales to column nations' manufacturing industries for seven largest Alliance's economies and China in 2019, again based on the Inter-Country Input-Output (ICIO) data from the OECD. As before, cell shades are indexed to share sizes; darker shades indicate higher bilateral FMR (more foreign market-dependent). Overall, the Alliance's economies' foreign market exposure with respect to China is high (higher than the bilateral foreign exposure between most Alliance's country pairs). Second, the global importance of the USA and China stand out from the rest, their respective columns are primarily shaded dark. However, the bilateral US-China asymmetry is less marked and reversed, since China's sales-side reliance on the US is 8.0% while that of the US on China's market is only 5.6%.

| (a) | USA | CAN | GER | GBR | FRA | ITA | JPN | CHN | ROW |
|---|---|---|---|---|---|---|---|---|---|
| USA |  | 5.4 | 1.8 | 1.0 | 0.7 | 0.8 | 2.1 | 9.9 | 13.0 |
| CAN | 32.5 |  | 2.1 | 1.5 | 0.9 | 0.9 | 2.0 | 11.8 | 21.1 |
| GER | 4.6 | 0.5 |  | 3.2 | 4.7 | 3.8 | 1.6 | 6.9 | 42.0 |
| GBR | 6.2 | 1.4 | 6.9 |  | 4.1 | 2.6 | 1.3 | 7.7 | 29.5 |
| FRA | 5.6 | 0.7 | 10.1 | 3.8 |  | 4.7 | 1.2 | 6.4 | 35.3 |
| ITA | 3.5 | 0.5 | 8.9 | 2.6 | 5.8 |  | 0.9 | 7.6 | 39.6 |
| JPN | 4.1 | 0.7 | 1.3 | 0.7 | 0.6 | 0.4 |  | 10.7 | 26.0 |
| CHN | 3.7 | 0.8 | 1.7 | 0.6 | 0.7 | 0.6 | 3.2 |  | 24.6 |

| (c) | USA | CAN | GER | GBR | FRA | ITA | JPN | CHN | ROW |
|---|---|---|---|---|---|---|---|---|---|
| USA |  | 3.2 | 1.0 | 0.8 | 0.7 | 0.4 | 1.3 | 5.6 | 9.6 |
| CAN | 31.9 |  | 0.8 | 1.3 | 0.6 | 0.3 | 1.7 | 10.8 | 9.3 |
| GER | 7.1 | 0.8 |  | 3.8 | 5.1 | 4.2 | 1.6 | 10.0 | 41.0 |
| GBR | 7.0 | 0.9 | 4.7 |  | 2.9 | 2.1 | 1.2 | 5.5 | 25.8 |
| FRA | 5.2 | 0.7 | 8.4 | 3.9 |  | 5.1 | 1.4 | 8.0 | 33.1 |
| ITA | 5.9 | 0.7 | 6.7 | 2.6 | 4.6 |  | 1.3 | 5.4 | 31.9 |
| JPN | 5.7 | 0.6 | 1.1 | 0.6 | 0.5 | 0.4 |  | 14.4 | 16.8 |
| CHN | 8.0 | 0.8 | 1.3 | 0.9 | 0.7 | 0.7 | 2.8 |  | 15.7 |

| (b) | USA | CAN | GER | GBR | FRA | ITA | JPN | CHN | ROW |
|---|---|---|---|---|---|---|---|---|---|
| USA |  | -1.4 | -0.6 | -0.5 | -0.2 | -0.2 | -1.8 | 6.0 | -3.9 |
| CAN | -1.1 |  | -0.2 | -0.8 | -0.2 | -0.1 | -1.6 | 6.1 | 2.0 |
| GER | 0.7 | -0.2 |  | -0.5 | -0.3 | -0.1 | -0.3 | 4.9 | 5.0 |
| GBR | 1.0 | 0.1 | 0.4 |  | -0.7 | -0.4 | -0.7 | 4.9 | 0.1 |
| FRA | 1.4 | 0.1 | 1.1 | -0.2 |  | -1.1 | -0.5 | 4.0 | 0.2 |
| ITA | 0.4 | -0.1 | 0.1 | -0.4 | -0.8 |  | -0.3 | 5.0 | 3.9 |
| JPN | 0.4 | 0.1 | 0.1 | 0.1 | 0.1 | 0.1 |  | 5.6 | 5.7 |
| CHN | -1.3 | -0.1 | -0.7 | -0.2 | -0.4 | -0.4 | -6.1 |  | -8.2 |

| (d) | USA | CAN | GER | GBR | FRA | ITA | JPN | CHN | ROW |
|---|---|---|---|---|---|---|---|---|---|
| USA |  | -0.5 | 0.1 | 0.1 | 0.1 | -0.1 | -0.3 | 3.8 | 1.4 |
| CAN | -17.6 |  | -0.1 | 0.1 | -0.1 | -0.2 | -1.1 | 8.3 | -0.2 |
| GER | -1.5 | 0.1 |  | 0.3 | 0.1 | -0.8 | 0.1 | 6.5 | 6.0 |
| GBR | -0.8 | 0.1 | 0.1 |  | -0.4 | -0.8 | -0.1 | 3.9 | 2.1 |
| FRA | -0.1 | 0.1 | 1.2 | -0.2 |  | -0.9 | -0.1 | 5.7 | 4.0 |
| ITA | 0.3 | 0.1 | 1.8 | 0.1 | -0.1 |  | 0.1 | 3.9 | 6.6 |
| JPN | -2.3 | -0.2 | 0.1 | -0.1 | -0.1 | -0.1 |  | 7.5 | 3.3 |
| CHN | -3.9 | -0.8 | -0.4 | -0.5 | -0.5 | -0.6 | -3.2 |  | -0.8 |

**Table 1: Panel (a): Foreign Input Reliance in 2019 (FIR, %); Panel (b): Change in Foreign Input Reliance between 2000 and 2019 (ppt); Panel (c): Foreign Market Reliance in 2019 (FMR, %); Panel (d): Change in Foreign Market Reliance between 2000 and 2019 (ppt). Source: Authors' computations based on Inter-Country Input-Output (ICIO) Tables http://oe.cd/icio. Notes: ROW denotes the Rest of the World.**

Finally, as for the input sourcing side, we also compute the change in the Foreign Market Reliance between 2000 and 2019. Table 1, panel (d) reports change in row nations' total input sales to column nations' manufacturing industries, 2019 vs. 2000. Dark-shaded cells indicate large FMR decreases or increases. Overall, panel (c) suggests that the Alliance's FMR has been further increasing with respect to China during the last two decades. These findings apply both to the input sourcing side as well to the sales side. Given that the foreign exposure is an inverse measure of the domestic industries' resilience and robustness with respect to aggregate shocks, our results imply that the increasing dependence on intermediate inputs from China and intermediate goods sales in China may magnify the Alliance's vulnerability.

## 2.2   Increasing aggregate Global Supply Chain ruptures

Because of a widespread production outsourcing, off-shoring, often insufficient investment in resilience and absence of robustness-promoting policies, many global production networks have become excessively complex and fragile (Baldwin and Freeman 2022). As a result, the GSCs of 2020s are efficient but brittle – vulnerable to breaking down in the face of a pandemic, a war or a natural disaster. These developments are important to understand, as the increasing fragility of GSCs may have implications for the vulnerability of critical sectors and essential services as well as implications for the entire society.

In the absence of systemic shocks to GSCs, the foreign input and output dependencies identified in Section 2.1 may not be critical. In reality, however, all production structures entail uncertainty, whereby sourcing inputs from abroad exposes domestic activity additionally to foreign shocks, making globally fragmented production structures more vulnerable than locally organised production processes. There are at least three propagation channels exposing the domestic activity to GSC shocks: the costs and effects of delinking; the propagation of micro shocks into macro shocks; and GSCs amplify the trade impact of macro shocks (Antras and Chor 2022).

Different metrics and indices have been developed to monitor and track ruptures of GSCs. The Global Supply Chain Pressure Index (GSCPI) is one of the most robust indices; it is being deployed by the Federal Reserve Bank of New York. GSCPI measures a common factor of several cross-country and global indicators of supply chain pressures (e.g., delays in shipments and delivery times and shipping costs after purging these from demand measured by new orders). As illustrated in Figure 1, the GSC pressures are at historically high levels since 2020, which is signalling an escalating frequency and scope of GSC ruptures.

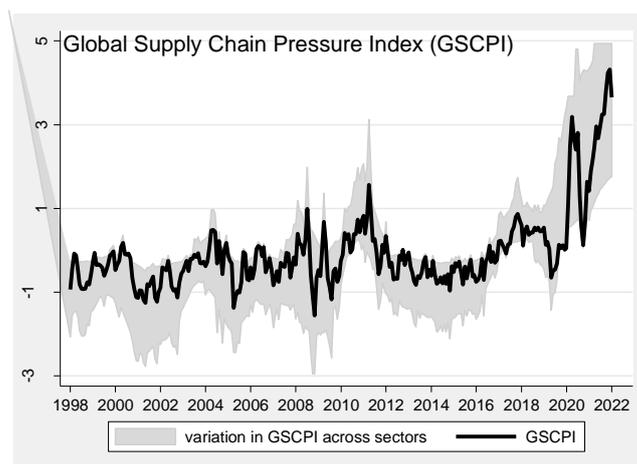

**Figure 1: Global Supply Chain Pressure Index (GSCPI) 1998-2022**
**Source: Computed based on www.newyorkfed.org/research/policy/gscpi**

The increasing frequency and intensity of GSC ruptures and concerns about the global disruption-caused losses have been synthesised and voiced in a White Paper of the World Economic Forum (2021):[8] "The increasing frequency of supply-driven disruptions – ranging from global pandemics and the climate crisis to cyber threats and geopolitical tensions – combined with an ever intensifying set of demand-driven disruptions – including the rise of new consumer channels, pent-up demand and a fragmented reopening of the global economy – will continue to destabilise global value chains."

## 3.0   PREVIOUS LITERATURE

Our study is related to two strands of inquiry that have evolved largely in parallel with some intersection only recently: the global supply (value) chain literature (e.g. Antras and Chor 2022) and literature on uncertainty by distinguishing explicitly between risk and ambiguity (see Ilut and Schneider 2023 for theoretical approaches, and Cascaldi-Garcia et al. 2023 for empirical issues). Our study contributes to the supply chain and uncertainty literature by considering ambiguity on the supply side of the economy, and by analysing optimal choices of ambiguity-averse firms facing uncertainty in input sourcing chains. To provide a context, we start the brief review of the relevant literature by evidencing the propagation of idiosyncratic and aggregate shocks through supply chains and the cascading disruptions at the aggregate level (e.g. Elliott and Golub 2022; Todo et al. 2023).

The empirical evidence reviewed by Elliott and Golub (2022) document the recent idiosyncratic shock propagation through supply networks. Disruptions to specific upstream firms are often found to cascade through input-output networks and cause widespread ruptures. In addition, regional disruptions, such as natural disasters, can have a contagion impact also in industries that are not concentrated, because it is common for many firms in a given industry to co-locate. Barrot and Sauvagnat (2016) and Boehm et al. (2019) analyse how economic shocks to suppliers due to natural disasters propagate to their customer firms because of input shortages. Carvalho et al. (2021) and Kashiwagi et al. (2021) found an upstream propagation from customers to their suppliers because of shortages of demand for inputs. Using data on natural disasters and investigating the consequences of firm-specific shocks, a strong evidence of shock propagation and supply chain fragility is identified. Shocks to suppliers propagate to sales of their customers; the effect is especially contagion when the disrupted supplier is producing hard-to-substitute relationship-specific inputs. Such shocks tend to further propagate to other (unaffected) suppliers of the customer firm. For example, the disruptions associated with the Great East Japan Earthquake have cascaded throughout the production network and had a negative shock on sales for firms with direct suppliers in disaster areas as well as propagated further downstream to the disaster area firms' indirect customers and upstream to the suppliers of firms hit by the earthquake. Hayakawa and Mukunoki (2021) confirmed the existence of upstream and downstream propagation from countries affected by the COVID-19 pandemic through global supply chains. Todo et al. (2023) use firm-level survey data for ten ASEAN countries and India to study the degree of the impacts of COVID-19 on supply chains. Firm-level evidence from East Asia confirms that firms with diversified customers and suppliers are more resilient, mitigating the damage from supply chain disruption through the substitution of upstream suppliers.

---

[8]   www.weforum.org/whitepapers/the-resiliency-compass-navigating-global-value-chain-disruption-in-an-age-of-uncertainty

In addition to firm- and supply-chain-specific effects of upstream ruptures, input-output linkages provide a channel for shock propagation through the economy magnifying micro shocks into macro fluctuations (Carvalho and Tahbaz-Salehi 2019). On the one hand, if an upstream supplier suffers a negative productivity shock, the resulting price rise worsens productivity for its downstream customers and thus propagates through the supply chain. On the other hand, a networked production in the presence of nominal rigidities can magnify the welfare costs of inflation, and change the impact of monetary policy. Horvath (2000) quantify the macroeconomic importance of idiosyncratic shocks by calibrating a multisector model under the assumption of uncorrelated disturbances. Horvath finds that idiosyncratic shocks, amplified by inter-firm interactions, account for approximately two-thirds of aggregate fluctuations. Foerster et al. (2011) show that what could appear to an econometrician as a common shock may instead be the result of endogenous co-movement generated by the equilibrium interactions between various industries in a production network. Overall, estimations from structural models confirm that industry-specific shocks amplified in this way can account for a nontrivial share of macroeconomic volatility.

The decision making under such shocks is studied in the uncertainty literature (Kumar et al. 2023), which can be classified in two types of frameworks: the Bayesian model (Bayes 1763) and the multiple priors model (Gilboa and Schmeidler 1989). In the expected utility framework, model uncertainty and uncertainty about events described by models are both treated as risk, that is, uncertainty with known odds. The ambiguity literature (Ilut and Schneider 2023) in contrast refers to uncertainty when the odds are not known. Among the many terrifying aspects of the COVID-19 pandemic was the realisation of how little we knew about what would happen. In such situations, ambiguity averse agents are not confident enough to assign probabilities to events about which they have little or no information.

Among models with risk-averse Bayesian agents, those specifying (and estimating) a full general equilibrium model with a production side and endogenous consumption are of a particular relevance for our study, as a general equilibrium framework allows one to investigate for example the impact of robustness on the aggregate macro dynamics. Hong et al. (2023) study shock events and the optimal firm response in terms of resilience investment incentives and willingness to pay for insurance against future shocks in a stochastic general-equilibrium model. Aggregate shocks follow a Poisson process and destroy capital stock, affect equilibrium prices, and reduce the welfare of households. Firm investment in shock mitigation and resilience is most effective when (i) investment at the firm level reduces the exposure of a firm's capital to the shock (e.g. sandbags to protect buildings), and (ii) investment at the aggregate level (that requires a collective action) reduces the conditional damage of a shock arrival and tail risk for all agents in the economy (e.g., an early warning system, and civil preparedness).

A number of findings from general equilibrium models with an endogenous investment in risk mitigation are directly relevant for understanding the uncertainty facing supply chains. First, unexpected arrivals of aggregate shocks have not only direct effects of capital destruction but also indirect effects due to learning that the world is riskier than anticipated. Second, in decentralised environments (such as global supply chains), firms do not internalise the benefits of the aggregate risk mitigation and hence underinvest in the aggregate risk mitigation. An optimal tax on capital to fund government investment in reducing aggregate tail risks can restore the first-best solution while still maintaining a balanced budget. The 'Pigouvian wedge' (between private and public risk evaluations (Riabov and Tsyvinski 2021) is incorporated in our modelling framework, whereas the learning and belief updating is left as a promising avenue for future research.

As regards uncertainty models with ambiguity, most of them consider an endowment economy (Lucas 1978) due to their complexity. Focusing on the consumption side of the economy while abstracting from production allows one to investigate resource allocation decisions of agents under uncertainty based on marginal utilities evaluated at the equilibrium consumption process while ignoring the deeper features that made them into equilibrium outcomes. Usually, an ambiguity-averse representative agent has robust preferences in the spirit of the robust control theory (Hansen and Sargent 2008). The decision maker acknowledges that her baseline model can only approximate the true environment. Rather than using a single model (as Bayesian agents do), a robust agent considers a set of alternatives to the baseline. Instead of seeking a decision rule which is optimal under a single (unknown) model, the decision maker wants decision rules that perform 'well enough' under all alternative models that she finds reasonable. Seeking a decision rule robust to these alternative models, the decision maker discovers a worst-case model to which she responds optimally (Ilut and Schneider 2023). In a robust control framework, an agent has subjective beliefs, and model-implied beliefs distort experienced probabilities by over-weighting states with adverse utility consequences, yielding expectations that are pessimistic relative to the observable state dynamics. The max-min theory of Gilboa and Schmeidler (1989) allows to determine the robust decision rule and the worst-case model as optimal equilibrium decisions.

The key insights from the ambiguity literature of modelling uncertainty in the supply chain context is that while both risk and ambiguity lower welfare, ambiguity can generate first order welfare effects of uncertainty. Models with ambiguity aversion capture agents' doubts about a model specification and a parameter uncertainty through belief sets by making agents less confident. These mechanisms formulated in ambiguity models (Ilut and Schneider 2023) are also adopted in our analysis. Following these studies, a robust decision rule and the worst-case model guide ambiguity averse agent's decisions in the modelling framework of the present study, where an ambiguity averse agent with a set of beliefs that differ in mean evaluates shock event probabilities as if the worst case mean is low; uncertainty does not matter only for higher moments.

Those uncertainty models that focus specifically on supply chains are of a particular relevance for the present study. Typically, these models consider upstream and downstream firm decisions in risky multi-tier production networks. Jiang et al. (2022) study how input sourcing decisions of a downstream manufacturer changes in the presence of uncertainty and what are implications on the supply chain's robustness. Kopytov et al. (2023) investigate how a supply chain's uncertainty affects firms' sourcing decisions and how individual firm decisions affect the aggregate macro economy. Whereas Jiang et al. (2022) simulate an optimal strategy to aggregate shocks that perish all suppliers in a particular region, Kopytov et al. (2023) simulate an optimal strategy to aggregate productivity shocks that hit all suppliers in a particular sector. Our framework is closer to Jiang et al. (2022) in terms of modelling shocks (reduced-form aggregate destructive events hitting one location), but closer to Kopytov et al. (2023) in terms of upstream and downstream interactions (firms can adjust their input sourcing decisions through both intensive and extensive margins at the beginning of the period, while once the production process has started only the intensive margin is active until the end of the period).

Jiang et al. (2022) study uncertainty's impact on the agents' decision making in a two-tier (upstream suppliers and a downstream manufacturer) vertical supply chain model with two regions that differ in the probability that an aggregate shock hits. Production is uncertain: in each period, one of the two regions might suffer an aggregate shock with an unknown arrival probability. In the uncertainty scenario, the conditional distribution of the aggregate shock is unknown. Each upstream supplier only produces one unit of a horizontally differentiated intermediate input, which has a constant cost paid before the production takes off. Importantly, the cost of production is paid irrespectively of the shock realisation. The fix cost turns out to be a key driver of supply chain's robustness in Jiang et al. (2022). Upstream suppliers sell the intermediate inputs to the downstream manufacturer, who produces final demand goods. Each upstream firm faces a location decision - to choose one of the two locations in every period where to set up production. Upstream suppliers maximise the continuation value - the value of locating in each region by solving a static problem. When maximising instantaneous profits, they do not take into account the survival probability of the entire supply chain. As a result, all upstream firms locate production in the region with the lowest probability that an aggregate shock hits, exposing in such a way the supply chain to an aggregate shock in this region. This individual suppliers' solution contrasts with the central planner's solution - a fraction larger than zero of input suppliers located in every region. Whereas individual suppliers maximise efficiency, the central planner maximises the supply chain's survival, it internalises the survival externality. These results resemble also in our study.

Kopytov et al. (2023) study endogenous production network formation under supply chain uncertainty in a multi-sector economy. At the beginning of a period, to produce each downstream firm must choose upstream suppliers, which specifies which intermediate inputs to use and how these inputs are to be combined. Hence, a downstream firm can adjust the sourcing importance of a supplier in the supply chain or drop that supplier altogether, and upstream sourcing decisions, when aggregated, may lead to changes in the supply network along both the intensive and extensive margins. The economy is subject to random sector-specific productivity shocks. After shock realisations, downstream manufacturers can only adjust how much they produce and the quantity of inputs that they use from the chosen suppliers (not the set of suppliers). Given that firms' beliefs about the distribution of sectoral productivities can influence their choice of upstream suppliers, the probability distribution of the shocks can affect the input-output structure of the economy. Firms are owned by a risk-averse representative household, implying that they inherit the household's aversion toward uncertainty. Firms compare profits across different states of the world using a stochastic discount factor. A competitive pressure between producers implies that the shocks, which affect production costs, are reflected in output prices. While a firm would generally prefer to source inputs from a more productive supplier, it might decide not to do so if this supplier also faces higher uncertainty. Such a supplier would sell at a lower price on average, but it is also more likely to suffer from a large negative productivity shock, in which case the price of its output would rise substantially. Downstream manufacturers take this possibility into account and balance concerns about the average productivity and robustness when choosing upstream suppliers (efficiency-resilience trade-off). This specific productivity-robustness trade-off mechanism studied in this paper is also present in our model.

Summing up, the existing literature shows that global supply networks with highly complex interdependencies among upstream and downstream firms can propagate shocks far beyond the sectors and locations where they originate and cause disruptions of global supply networks. This vulnerability is particularly acute when multiple suppliers from different downstream chains rely on one/few firms which further upstream in the supply chain are unknown to the downstream manufacturer. In addition, a spatial colocation of firms within an industry can trigger cascading raptures by making them vulnerable to an aggregate shock. Uncertainty models offer insights of a decision making under risk and ambiguity and how the information precision affects the optimal agent's choices. Our contribution to the existing literature is two-fold. First, we apply the multiple-priors framework to a supply chain facing not only risk, but also ambiguity. Second, we study how ambiguity-averse decision makers can practically enhance resilience and ensure a chain-level robustness. Differently from the previous literature, our focus is on the question if, under what uncertainty environment and how much resources in terms of a foregone efficiency would a decision maker be willing to divert to an increased robustness.

# 4.0 MODELLING FRAMEWORK AND COUNTERFACTUAL SIMULATIONS

In order to investigate the relationship between the upstream supplier diversification and the chain-level robustness (survival probability), we augment a parsimonious supply chain model of Antras and de Gortari (2020) with uncertainty as in Jiang et al. (2022) and Kopytov et al. (2023) and simulate alternative upstream sourcing strategies in presence of information incompleteness and uncertainty of aggregate shocks. The key characteristics and channels of adjustment relevant to the analysis along with the main intuition of the modelling framework are explained in this section.[9] Using the model, we will simulate alternative information incompleteness scenarios, each of which is investigated under three shock realisation regimes. The main interest of our simulations is on targeted government actions that can improve the chain-level robustness. We will show that in the firms' pursuit of efficiency, a supply chain can become vulnerable to aggregate shocks. On the other hand, a supply chain with greater robustness to shocks has to sacrifice efficiency during normal times (in absence of shocks).

## 4.1 Baseline model

Time is discrete and infinite for tractability. The world economy consists of $J \geq 1$ locations, which are populated by representative consumers which derive utility from consuming differentiated varieties of a single sector good. Preferences across varieties are CES. The location-$i$ composite good, which is used both for the final consumption, as well as an intermediate input in production process by firms, is a CES aggregate over the set of varieties on the unit interval:

$$Q_i = \left( \int q_i(\omega)^{(\sigma-1)/\sigma} d\omega \right)^{\sigma/(\sigma-1)}$$

(1)

where $q_i(\omega)$ denotes the quantity of variety $\omega$ that is ultimately purchased in location $i$, sourced from the lowest-cost source locations. Preferences in (1) exhibit the love of variety, capturing the extent to which an additional product variety generates benefits in either final or intermediate consumption – the marginal taste for an additional variety. The love of variety is modelled in both the final and intermediate consumption as the marginal gain in proportional terms, from spreading a constant total amount of consumption across one additional variety. The love of variety is a force driving a supplier diversification.

Production technology is vertically organised. Downstream producers sourcing from upstream suppliers along the supply chain produce a composite good in $N$ production tiers that need to be performed sequentially. For the sake of tractability, when introducing uncertainty, we assume two tiers $N = 2$ in the counterfactual analysis, though the framework is more general and extends to any number of tiers straightforwardly (see Antras and de Gortari 2020). Firms in the upstream tier $n = 1$ only use local labour, $l$. Implicitly, labour consists of two parts: raw labour and human capital (which in reality may differently be hit by shocks), we model them together as 'labour'. Firms in the second tier combine the local labour with intermediate inputs produced in any location in the first tier, $y^1$. Both inputs – labour and intermediate components – are required in the tier-two production, no output can be produced with one input only. If either labour or at least one source of intermediate inputs (or both) are not supplied, no production can take place in this location. Formally, the respective tier 1 and tier 2 production technologies can be expressed as follows:

$$y_i^1(\omega) = z_i^1(\omega) l_i^1(\omega)$$

(2)

$$y_i^2(\omega) = \left( z_i^2(\omega) l_i^2(\omega) \right)^{\alpha_2} \left( y_i^1(\omega) \right)^{1-\alpha_2}$$

---

[9] For formal derivations, proofs of the underlying supply chain model, see Antras and de Gortari (2020).

(3)

where $\alpha_2 \in (0,1)$ denotes the labour share in tier-2 production, and $z_i^n(\omega)$ is labour productivity at tier $n$ in location $i$. All tiers of the supply chain are characterised by a perfect competition among firms. The optimal input sourcing location $l(n) \in \{1,...,J\}$ at the different production tiers $n \in \{1,2\}$ of the supply chain is determined by a cost minimisation of downstream firms.

Final goods can be produced using different sets of intermediate inputs that can be sourced from any location. Kopytov et al. (2023) label these firm decisions as techniques. Slightly differently in our framework, a technique is a production function that specifies from which locations and suppliers intermediate inputs to source. Techniques can also differ in terms of productivity. When choosing a production technique, a firm can adjust the importance of a supplier e.g. by including or dropping that supplier. Production techniques are chosen and the beginning of the period, firms have to produce with the selected input suppliers. These decisions, when aggregated, lead to changes in the supply network along both the intensive and extensive margins, which have implications to the supply chain's robustness. Firms can adjust the input sourcing path at the end of the period.

Locations can differ in three key aspects: (i) their size, as reflected in the measure $L_i$ of labour force available for production in each location $i$; labour is inelastically supplied and commands wage $w_i$, (ii) their geography, as captured by a $J \times J$ matrix of iceberg trade cost $\tau_{ij} \geq 1$, and (iii) their technological efficiency, as determined by the labour productivity terms $z_i^n(\omega)$. Following Eaton and Kortum (2002), we assume that $z_i^n(\omega)$ is drawn independently (across tiers) from a Fréchet distribution with a cumulative distribution function $F_i^n(z) = \exp\{-T_i^n z^{-\theta}\}$.[10]

The downstream firm's problem in the vertically organised supply chain consists of choosing the least-cost path of production to manufacture a final composite good's variety $\omega$ and deliver to consumers in location $j$. Given equations (2) and (3), this amounts to choosing input sourcing locations $l_j^1$ and $l_j^2$ to minimise

$$c(l_j^1, l_j^2) = \left(\frac{w_{l_j^2}}{z_i^2(\omega)}\right)^{\alpha_2} \left(\frac{\tau_{l_j^1}\tau_{l_j^2}w_{l_j^1}}{z_i^1(\omega)}\right)^{1-\alpha_2}$$

(4)

Given the Fréchet distribution on the labour productivity $z_i^1(\omega)$ and $z_i^2(\omega)$ yields a distribution for the equilibrium marginal cost of production of the supply chain, and facilitates a description of the general equilibrium solution. Note that due to non-linearities, the minimum cost (4) associated with a given supply chain's production path cannot be characterised by an analytically tractable solution. To gain tractability and allow the characterisation of some of the equilibrium features, we follow Antràs and de Gortari (2020) and treat the overall (i.e., chain-level) unit cost of production of a supply chain flowing through a sequence of locations as a draw from a Fréchet random variable with a location parameter that is a function of the states of technology and wage levels of all locations involved in the supply chain, as well as of the trade costs incurred in that supply chain.[11] Consider a given production path $\boldsymbol{l} = \{l_j^1, l_j^2\} \in J^2$, where $J$ denotes the set of locations in the world. The given supply chain's production cost is a function of trade costs, composite factor costs and the state of technology of the various locations involved in the chain. Nevertheless, two supply chains flowing across the same locations in the exactly same order may not achieve the same overall productivity due to idiosyncratic shocks.

Following Antràs and de Gortari (2020) and given the cost function (4), the overall productivity of a given supply chain $\boldsymbol{l} = \{l_j^1, l_j^2\}$ is characterised by

$$\Pr\left(\left(z_i^1(\omega)\right)^{1-\alpha_2}\left(z_i^2(\omega)\right)^{\alpha_2} \leq z\right) = \exp\left\{-z^{-\theta}\left(T_{l^1}^1\right)^{1-\alpha_2}\left(T_{l^2}^2\right)^{\alpha_2}\right\}$$

---

[10] The GSC complexity and opaqueness, and incomplete information regarding upstream suppliers discussed in section 2 provide a justification for the Fréchet distribution.

[11] The Fréchet random variable accounts for idiosyncrasies characterising real-world supply chains. For discussion, see section 2.

(6)

which is equivalent to assuming that $\left(z_i^1(\omega)\right)^{1-\alpha_2}\left(z_i^2(\omega)\right)^{\alpha_2}$ is distributed Fréchet with a shape parameter $\theta$, and a location parameter that is a function of the states of technology in all locations in the chain, as captured by $\left(T_{l^1}^1\right)^{1-\alpha_2}\left(T_{l^2}^2\right)^{\alpha_2}$. A direct implication of this assumption is that the unit cost associated with serving consumers in a given location $j$ via a given supply chain $\boldsymbol{l}$ is also distributed Fréchet, which then allows to characterise equilibrium prices and the relative prevalence of different supply chains.

The share of location $j$'s income spent on the final composite good produced under a particular supply chain's input sourcing production path $\boldsymbol{l} \in J^2$ is given by

$$\pi_{lj} = \frac{\left(\left(T_{l^1}^1\right)^{\alpha_n}\left((w_{l^1})^{\alpha_n}\tau_{l^1 l^2}\right)^{-\theta}\right)^{1-\alpha_2} \times \left(T_{l^2}^2\right)^{\alpha_2}\left((w_{l^2})^{\alpha_2}\tau_{l^2 j}\right)^{-\theta}}{\sum_{l\in J^2}\left(\left(T_{l^1}^1\right)^{\alpha_n}\left((w_{l^1})^{\alpha_n}\tau_{l^1 l^2}\right)^{-\theta}\right)^{1-\alpha_2} \times \left(T_{l^2}^2\right)^{\alpha_2}\left((w_{l^2})^{\alpha_2}\tau_{l^2 j}\right)^{-\theta}}$$

(7)

and the exact ideal price index $P_j$ in location $j$ is given by

$$P_j = \kappa \left(\sum_{l \in J^2} \left(\left(T_{l^1}^1\right)^{\alpha_n}\left((w_{l^1})^{\alpha_n}\tau_{l^1 l^2}\right)^{-\theta}\right)^{1-\alpha_2} \times \left(T_{l^2}^2\right)^{\alpha_2}\left((w_{l^2})^{\alpha_2}\tau_{l^2 j}\right)^{-\theta}\right)^{-1/\theta}$$

(8)

where $\kappa$ is a constant that depends only on $\sigma$ and $\theta$. Note that for the price index to be well defined, we need to impose $\sigma - 1 < \theta$.

Equations (7) and (8) have a number of implications for the supply chain robustness. On the one hand, firms producing in locations with higher states of technology, $T_i$, or lower labour costs, $w_i$,[12] will tend to participate disproportionately in the intermediate input supply, leading to upstream supplier concentration. On the other hand, firms producing in locations with higher transportation costs, $\tau_{l^1 l^2}$, are penalised regarding their participation in global supply chains, leading to fewer upstream suppliers in the supply chain. The compounding effect on the supply chain's robustness depends on the exact model's parameterision.

To solve for equilibrium wages, we note that for any supply chain, tier-$n$ value added (labour income in our model) accounts for share $\alpha_n \beta_n$ of the value of the final demand good emanating from that supply chain. Further, the total spending in any location $j$ is given by $w_j L_j$, and the share of that spending by location $j$ flowing to supply chains in which location $i$ firms participate in tier $n$ is given by $Pr(\Lambda_i^n, j) = \sum_{l \in \Lambda_i^n} \pi_{lj}$ where $\Lambda_i^n = \{l \in J^N | l^n = i\}$ and $\pi_{lj}$ is given in equation (7). It thus follows that the equilibrium wage vector in location $i$ is determined by the solution of the following system of equations

$$w_i L_i = \sum_{j \in J} \sum_{n \in N} \alpha_n \beta_n \times Pr(\Lambda_i^n, j) \times w_j L_j$$

(9)

The system of equations is nonlinear, because $Pr(\Lambda_i^n, j)$ is a nonlinear function of wages themselves, and of the output price vector $\boldsymbol{P}$, which is in turn a function of the vector of wages $\boldsymbol{w}$. In a supply chain with only one production tier, $N = 1$, equation (9) would imply that $\alpha_n \beta_n = 1$ and $Pr(\Lambda_i^n, j) = \pi_{ij} = \left(\tau_{ij} c_i\right)^{-\theta} T_i^1 / \sum_k \left(\tau_{kj} c_k k\right)^{-\theta} T_k^1$.

The produced composite good is both consumed in the final demand and used as an intermediate input in production process. First, consider the consumers' final demand. For final demand goods to flow from source location $i$ to destination location $j$, it must be the case that location $i$ is in position $N$ in a supply chain serving customers in location $j$. Defining the set of supply chains flowing through location $i$ at position $n$ by $\Lambda_i^n \in J^{N-1}$, the overall share of spending in location $j$ on goods value added in location $i$ (i.e., in supply chains in which location $i$ produces tier $N$) can be expressed as

$$\pi_{ij}^F = \frac{\sum_{l \in \Lambda_i^N} \prod_{n=1}^{N-1}\left(\left(T_{l^n}^n\right)^{\alpha_n}\left((c_{l^n})^{\alpha_n}\tau_{l^n l^{n+1}}\right)^{-\theta}\right)^{\beta_n} \times \left(T_i^N\right)^{\alpha_N}\left((c_i)^{\alpha_N}\tau_{ij}\right)^{-\theta}}{\sum_{l \in J^N} \prod_{n=1}^{N-1}\left(\left(T_{l^n}^n\right)^{\alpha_n}\left((c_{l^n})^{\alpha_n}\tau_{l^n l^{n+1}}\right)^{-\theta}\right)^{\beta_n} \times \left(T_{l^N}^N\right)^{\alpha_N}\left((c_{l^N})^{\alpha_N}\tau_{l^n j}\right)^{-\theta}}$$

(10)

where the composite cost, $c_i$, in location $i$ is captured by a Cobb-Douglas aggregator, $c_i = (w_i)^\gamma(P_i)^{1-\gamma}$ with $P_i$ being the ideal price index associated with preferences. Equation (10) implies that the final demand trade flows between any two locations $i$ and $j$ are then given by $\pi_{ij}^F \times w_j L_j$.

---

[12] In Table 1 (section 2), high states of technology are e.g. in US and Germany, whereas low labour costs e.g. in China. In both cases, the participation in GSCs is disproportionately high, as predicted by the model.

Analogously, the intermediate input demand can be derived. Based on the supply-chain's trade shares in (7), intermediate input flows between any two locations $i$ and $j$ can be expressed as

$$\pi_{ij}^T = \frac{\prod_{n=1}^{N-1}\left(\left(T_{l^n}^n\right)^{\alpha_n}\left((c_{l^n})^{\alpha_n}\tau_{l^n l^{n+1}}\right)^{-\theta}\right)^{\beta_n} \times \left(T_{l^N}^N\right)^{\alpha_N}\left((c_{l^N})^{\alpha_N}\tau_{l^n j}\right)^{-\theta}}{\sum_{l \in J^N} \prod_{n=1}^{N-1}\left(\left(T_{l^n}^n\right)^{\alpha_n}\left((c_{l^n})^{\alpha_n}\tau_{l^n l^{n+1}}\right)^{-\theta}\right)^{\beta_n} \times \left(T_{l^N}^N\right)^{\alpha_N}\left((c_{l^N})^{\alpha_N}\tau_{l^n j}\right)^{-\theta}}$$

(11)

Note that intermediate input demand (11) is more involved than final goods demand (10), as to solve for $\pi_{ij}^T$ one needs to take into account both vertical trade between two contiguous tiers as well as intermediate input trade flows across locations associated with the use of the bundle of inputs at each tier.

### 4.2 Partial information and uncertainty[13]

In absence of uncertainty, the optimal input sourcing paths of downstream firms are selected by weighing the relative states of technology, $T_i$, labour costs, $w_i$, and transportation costs, $\tau_{l^1 l^2}$, between locations at every tier. A supply chain's diversification is implied by the love of variety aggregator in (1), according to which everything else equal sourcing from more suppliers (each producing a distinct variety) generates higher marginal gain. The production process however is uncertain (Kumar et al. 2023), for example, a location may be hit by an aggregate shock in any period. Specifically, in every period $t$, one of the locations may suffer an aggregate shock with arrival probability $\eta$, there is no aggregate shock in a location with probability $1 - \eta$. The information about the shock realisation is incomplete to firms: firms do not know ex-ante if there will be a shock in the next period and if so in which location the shock will realise. Differently from the previous literature, where often shock intensity is the main variable of interest, the shock intensity parameters are constant and known to agents in our framework. The aggregate shock hitting a location is assumed to lead to a loss of the entire local labour force for the duration of the 'shock' event – one period. As a rationale for this assumption, we may think about the stay-at-home orders and other lockdown regulations limiting the labour force participation during the COVID-19 pandemic. The modelled shocks are aggregate in sense that they affect the entire labour force in the location. The shocks are local in sense only one location can be hit by the shock in every period. The entire labour force fully reverts to the pre-shock dynamics once the location switches to a 'normal' state. I.e., after the shock, the location looks stochastically the same as it did before. In particular, there are no long-run scarring effects, e.g., on the location's human capital. In this sense, the model is optimistic.

The economy in every location transitions stochastically between shock episodes and "normal times". Firms optimally solve their production problem, whose solution depends on both the current state of the economy and on the current information about the unobserved shock realisation parameters. Following Acharya et al. (2023), we consider the state of economy in every location to be either in a "normal" regime or in a "shock" regime, and denote the state as $s \in \{0,1\}$. We assume that the local economy switches between these states with the following transition probabilities:[14]

---
[13] Uncertainty in the baseline model is introduced following Jiang et al. (2022) and Kopytov et al. (2023).

$$\Pr(s_{t+dt} = 1 | s_t = 0) = \eta dt \tag{13}$$

$$\Pr(s_{t+dt} = 0 | s_t = 0) = 1 - \eta dt \tag{14}$$

$$\Pr(s_{t+dt} = 0 | s_t = 1) = \lambda dt \tag{15}$$

$$\Pr(s_{t+dt} = 1 | s_t = 1) = 1 - \lambda dt \tag{16}$$

where $\eta$ and $\lambda$ are the probabilities per unit time of switching from one state to the other. Specifically, $\eta$ is the intensity of switching from state 0 ("normal") to state 1 ("shock") and $\lambda$ is the intensity of switching from 1 to 0. These parameters are unobservable to firms.

For the sake of the model's tractability in the presence of uncertainty, we assume two locations, $L = 2$, in the counterfactual analysis. Conditional on the shock realisation and before the input sourcing and production decisions are made, the entire labour force in the location hit by the shock experiences an exogenous rupture with probability $\zeta$ or $1 - \zeta$, respectively. An aggregate shock to a location 'locks down' the entire local labour force, which reduces the number of active firms participating in the supply chain and hence the number of final good's varieties delivered to consumers of final demand goods and downstream manufacturers. The survival of the entire supply chain is conditional on at least one firm (from all locations together) producing at each production tier.

To gain intuition, it is useful to consider an aggregate shock hitting one of the two locations. Only 'fully diversified' supply chains, i.e. those sourcing intermediate inputs from both locations and performing all tiers in both locations, as well as 'purely-local' supply chains in the other location would survive. Such a 'purely-local' supply chain performs all tiers in a given location $j$ to serve customers solely in the same location $j$. To the effect, denote the local supply and local absorption by $j = (j, j, \ldots, j)$. Plugging price index in (8) into equation (7) and simplifying terms yields a closed-form expression for gains from participation in a supply chain with $N$ sequential production tiers:

$$\frac{w_j}{P_j} = \left(\kappa(\tau_{jj})^{\sum_{n=1}^{N} \beta_n}\right)^{-1} \left(\frac{\prod_{n=1}^{N}(T_j^n)^{\alpha_n \beta_n}}{\pi_{jj}}\right)^{1/\theta} \tag{12}$$

where $\pi_{jj}$ is the share of spending on goods that are produced entirely through local supply chains. Note that no such closed-form expression for 'fully diversified' supply chains can be derived analytically, as one needs to take into account not only the vertical trade between two contiguous tiers but also the entire intermediate input trade between locations associated with the use of the bundle of inputs at each tier.

### 4.3 Simulation scenarios

The conceptual framework introduced in the previous sections contains an efficiency-resilience trade-off and several channels of adjustment (intensive and extensive margins). Specifically, firms are owned by a representative household in each location, they compare profits across different states of the world. As such firms inherit the household's attitude toward uncertainty. Consequently, while a firm would generally prefer to purchase intermediate inputs from a location with higher state of technology, lower labour cost, and lower bilateral transportation costs, it might decide not to do so if this location faces also higher uncertainty. Firms in such a location would sell at lower price on average, but it is also more likely to suffer from an aggregate shock, in which case the intermediate and final goods production would perish in this location. Potential final and intermediate demand customers take this possibility into account and balance concerns about the average productivity and robustness when choosing a production technique (sourcing path across locations).

---
[14] For the sake of brevity and clarity, in introducing shock events we omit subscripts referring to locations.

How exactly is an aggregate supply chain's robustness (survival probability) related to input sourcing decisions at the firm level? How can a central planner (i.e. government) simultaneously ensure resilience and a continued flow of essential goods under a systemic shock? We attempt to answer these questions in this section by formulating and simulating two uncertainty scenarios each of which is studied under three different shock realisations. To study firm sourcing decisions on the chain-level robustness, in counterfactual simulations we presume that $\zeta \gg 1/2$, implying that one location (East) is riskier than the other location (South). In the following period, $t+1$, the production processes takes place by the surviving firms along the chosen supply chain. The three shock realisation possibilities with six supply chain rupture outcomes in total are as follows: (i) there is no aggregate shock (with probability $1-\eta$); (ii) the aggregate shock hits East in period 10 and all firms in East perish (with probability $\eta * \zeta$);[15] (iii) the aggregate shock hits South in period 10 and all producers in South perish (with probability $\eta * (1-\zeta)$). The three possible shock realisations are summarised in the columns of Table 2. The alternative location decisions of upstream suppliers will allow us to investigate individual firm sourcing decision impact on the aggregate supply chain's robustness.

|  | Risky information environment | Ambiguous information environment |
| --- | --- | --- |
| No aggregate shock | Known shock distribution, no aggregate shock | Unknown shock distribution, no aggregate shock |
| Aggregate shock in East | Known shock distribution, shock realisation in East | Unknown shock distribution, shock realisation in East |
| Aggregate shock in South | Known shock distribution, shock realisation in South | Unknown shock distribution, shock realisation in South |

**Table 2: Simulation scenarios: key assumptions about shock realisations (rows) and uncertainty - information incompleteness (columns)**

The information about the spatial realisation of a shock event is incomplete to firms: firms do not know ex-ante which location (and if any) will be hit by an aggregate shock. In counterfactual simulations, we study how the incompleteness of information about the shock realisation affects individual supplier sourcing decisions and the chain-level robustness. In line with the Knightian uncertainty, there are two categories of imperfectly predictable events, which involve a differentiated agents' behaviour: risky events and ambiguous events.[16] (i) In the risk scenario, we assume that the distribution of the conditional probability that the aggregate shock realises in the particular location is known, the shock probability follows some a priori known distribution (in the baseline simulations we assume that the distribution is uniform, alternative distributions are explored in sensitivity analysis). (ii) In the ambiguity scenario, we assume the shock exhibits a bounded uncertainty and the distribution of the conditional probability that the aggregate shock event realises in the location is unknown. Agents know that a shock can occur but they do not know even the underlying distribution on which to form their beliefs about the uncertain shock event. Ambiguity averse agents are not confident enough to assign probabilities to events about which they have little information. These differences in the conditional probability that an aggregate shock hits a particular location are summarised in rows of Table 2. In counterfactual simulations we will see that the information incompleteness and asymmetry poses a serious challenge to agents in finding an optimal (both efficient and robust) supplier path.

---

[15] Note that the period of the shock does not change the qualitative outcomes of results.
[16] Some literature e.g. Jiang et al. (2022) refers to them as risk and uncertainty.

In a further set of simulations, we investigate the role that a government (central planner) can play to mitigate the supply chain's vulnerability to shock events and enhance resilience e.g. by a chain-level coordination of upstream sourcing decisions. To analyse an optimal upstream supplier sourcing diversification pattern that accounts for both the supply chain's efficiency but also robustness, we simulate the above described two information incompleteness scenarios under three shock realisation scenarios in a central planner's optimisation problem. Key modelling assumptions under each simulation scenario remain the same as above and as summarised in Table 2.

### 4.4 Simulation results

Baseline simulation results when individual firms optimise their production techniques without coordinating mutually their upstream sourcing paths are reported in Figure 4. The number of upstream suppliers in each location are presented as dashed grey line for East and squared black line for South, respectively, the total number of intermediate input suppliers as a solid red line. The latter is the one the final goods manufacturer cares the most.

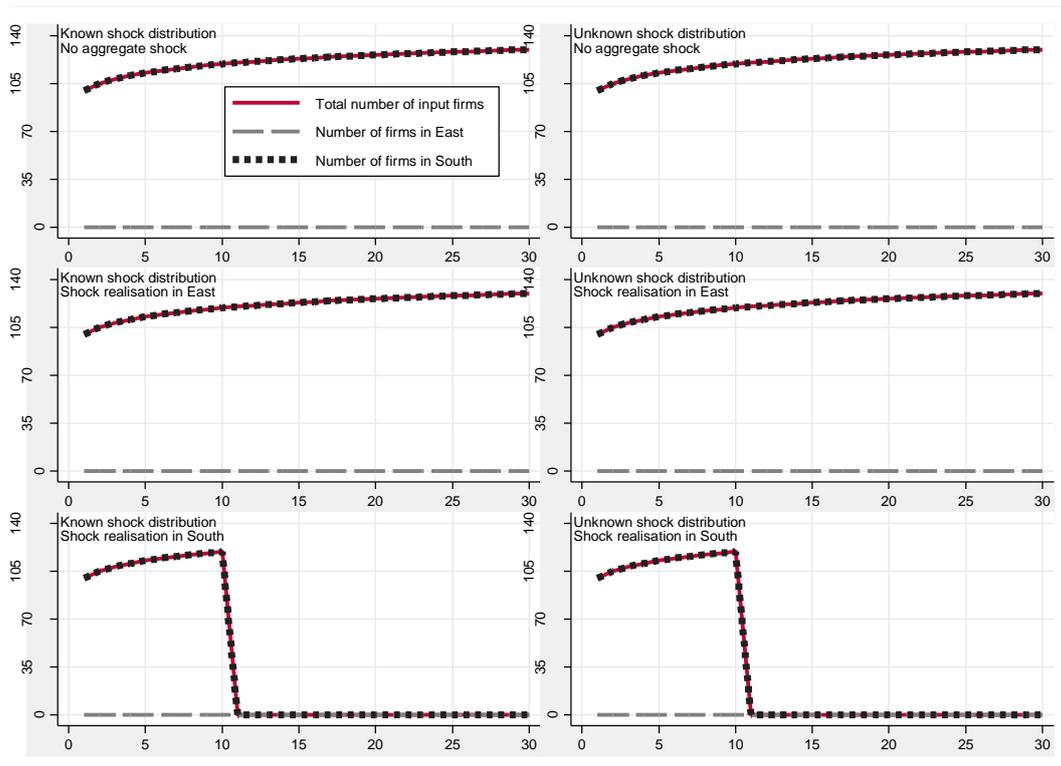

**Figure 4: Simulation results: supply chain's survival under various shock realisation and information incompleteness scenarios, individual supplier optimisation.**
**Notes: Y-axis measures the number of intermediate input suppliers; X-axis time periods. Risky environment is on left; ambiguous environment in right panels.**

First, note that given our assumption of lower shock probability in South,[17] the continuation value for each downstream manufacturer sourcing from suppliers in South is always larger than the expected value of sourcing from suppliers in East. This implies that South is the preferred input source for every downstream firm, they all contract suppliers in South. While every downstream manufacturer maximises its efficiency, collectively, they expose the entire supply chain to an aggregate shock in South. Recall that production techniques (input sourcing paths) are chosen at the beginning of the period, firms have to produce with the selected input suppliers in the current period. Indeed, this can be seen in the two bottom panels, where after an aggregated shock in period 10, the number of upstream suppliers in South drops to zero (squared black line). Given that all suppliers have perished in South, the entire intermediate input supply to downstream manufacturers drops to zero (solid red line). Because no final good manufacturing can take place without all input components (labour and at least one supplier of intermediate goods), the entire supply chain is disrupted. Hence, such a supply chain is not robust to aggregate shocks in a pure market-driven driven environment where input sourcing decisions of individual firms do not internalise the network externality. Recent examples of such upstream-concentrated supply chains include European energy extraction chains concentrating in Russia until 2022 and the ICT component suppliers' concentration in China.

---
[17] Note that instead assuming, for example, lower costs in South would yield qualitatively equal results.

As regards the role of uncertainty and information precision on the supplier diversification, the results under both risky and ambiguous information environments are identically non-robust, when the input source diversification is an outcome of individual firms trading their current period's efficiency off the resilience. Given that ambiguity and risk both are concepts of uncertainty, uncertainty of either flavour lowers welfare, both leading to a sub-optimal behaviour and uncertainty premium. Uncertainty drives a wedge between the expected realisations and the experienced realisations of shock events. Information imprecision creates a bias in the individually optimal suppliers' diversification strategy compared to a full-information baseline (not shown in Figure).

Such input-output paths of a supply chain's vulnerability are efficient in 'normal times' but are not robust under aggregate shocks. Given the currently less and less predictable global security environment, ensuring robust supply chains in the Alliance may be particularly important, for example, in critical and strategic sectors. As next, we explore if a central planner (government) can take actions to increase the supply chain's robustness with respect to aggregate shocks? For example, can a coordinated upstream supplier diversification strategy mitigate the systemic vulnerability to aggregate shocks? To answer these questions, as next, we simulate the same two information incompleteness scenarios under three shock realisation scenarios in a central planner's optimisation problem. The efficiency-robustness trade-off needs is present also under a central planner's optimisation, however, the perspective is different. Whereas individual downstream firms maximise their own value function, a chain-level optimisation considers the entire supply chain's continuation value, it internalises the network externality. Simulation results for the supplier diversification under a central planner's optimisation are reported in Figure 5.

In contrast to individual downstream manufacturers, the central planner considers in its optimisation also the entire supply chain's resilience. The central planner decides to source intermediate inputs also from East, to insure the supply chain's robustness against an aggregate shock in South. It internalises the supply chain's survival probability and diversifies the input sourcing. This can be seen in Figure 5, where the dashed grey lines (number of firms in East) are above zero in all scenarios – the number of intermediate input suppliers is larger than zero in all locations, including East. In contrast to individual optimisation outcomes reported in Figure 4, where the optimal individual firm strategy leads to a corner solution, the results shown in Figure 5 report internal solutions (the total number of surviving firms is greater than zero) in all scenarios.

Further, we can observe how information precision affects agents' choices and the aggregate preparedness to shock events; and how the optimal efficiency-resilience decisions change depending on the information incompleteness. In a risky information environment, Bayesian agents have imprecise information about the possible shock event – they only have information about the probability distribution of shocks (left panels of Figure 5). In contrast, in an ambiguous information environment agents are not confident enough to assign probabilities to shock events about which they have little information (right panels of Figure 5). When the uncertainty is very large and ambiguity averse agents realise that important aspects of the uncertainty may be unknown, uncertainty about probability assessments instead is described by the size of the belief set of agents, similarly to how model uncertainty is handled in a related literature on robust statistics (Maronna et al. 2018). When ambiguity averse agents make decisions, they act as if they are relatively pessimistic about expectations, and more so the larger their belief set (the lower is information precision).

In the left panels of Figure 5, a risk-averse central planner exhibits an allocation behaviour that takes into account all the sources of risk resulting in a desire for a supplier diversification. The diversification is obtained as in a traditional expected utility maximisation problem by either increasing the variance or the risk aversion. The central planner faces a trade-off between instantaneous profits (what also the individual firms maximise) and the probability of aggregate survival (robustness). The left panels in Figure 5 also suggest that the optimal diversification of intermediate suppliers and hence the input source diversification depends on the number of suppliers. If there are two input suppliers in total (one per location), the optimal production technique (input supply paths) of a risk-averse planner is to source equally from both firms. If the number of suppliers is large enough in all locations, depending on the slope of the uncertainty-reward curve, a fraction larger than one half of input suppliers will be chosen from South. Note that this is not the case under individual downstream firm optimisation, where a final good producer does not take into account input sourcing decisions of other firms.

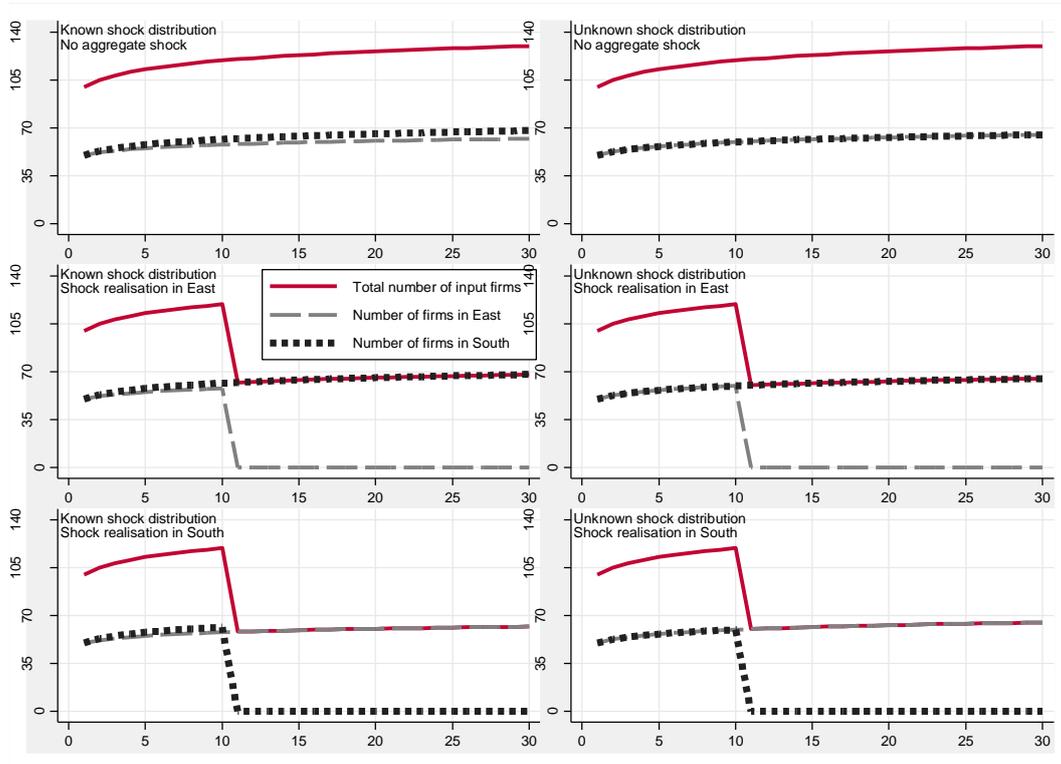

**Figure 5: Simulation results: supply chain's survival under various shock realisation and information incompleteness scenarios, central planner's optimisation.**
**Notes: Y-axis measures the number of intermediate input suppliers; X-axis time periods. Risky environment is on left; ambiguous environment in right panels.**

In contrast, an ambiguity-averse central planner's optimal production technique is independent of the number of suppliers in every location (right panels in Figure 5). The flows and costs are identical, therefore, the expected continuation value is independent of the true shock realisation. It does not matter how much more uncertainty faces East relative to South, there is a level of ambiguity for which the central planner sources half intermediate inputs from suppliers in East and half from South. A robustly optimal diversification of intermediate input suppliers is uniform across locations; in our example, half of the intermediate inputs sourcing from suppliers in East and half from firms in South. Such a robust strategy implies that in the presence of an aggregate shock – independently where it realises – at least half of firms would survive. According to the right panels of Figure 5, a robust supply chain is one in which the survival probability is maximised and where production will continue even under the most demanding circumstances.

A robustly optimal allocation is the optimal strategy, when the supply chain's survival is crucial, and the decision maker is not confident enough to assign probabilities to events about which it has little information. When facing ambiguity, the agent choses to be prepared for the worst. The simulation results presented in the right panels of Figure 5 show outcomes of input sourcing strategies that reduce the differences over all possible states of shock realisations. The central planner's robustly optimal strategy guaranteeing the survival of the supply chain is comparable to insights from behavioural theories, where in many cases agents tend to choose the robust survival maximising strategy (Ilut and Schneider 2022).

In summary, we may conclude that under an individual optimisation, the equilibrium firm choices are independent of the incompleteness of information. Individual firms choose a solution that is not robust at the chain-level, i.e. all supply chains rely on upstream suppliers in South, irrespectively of whether firms are facing risk or ambiguity. The entire supply chain is vulnerable to an aggregate shock to South (bottom panels in Figure 4). The optimal supplier diversification is rather different under a central planner's optimisation (Figure 5), which tends to result in internal solutions – a proportion larger than zero of intermediate input suppliers are chosen both from South and East in all scenarios. When the central planner faces a risky environment, the optimal diversification of upstream suppliers is an increasing function of the number of suppliers. Because East is more uncertain than South, a larger fraction of upstream suppliers are from South (left panels in Figure 5). When the information incompleteness points to ambiguity, the optimal supply chain's diversification strategy follows a robust decision-making with an equal and fixed fraction of intermediate inputs sourced from every location (right panels in Figure 5).

# 5.0 CONCLUSIONS

The production processes are increasingly fragmented across international borders. In the same time, the landscape of hybrid threats is expanding including new forms such as the weaponisation of supply chains. The recently experienced aggregated shocks such as the Covid-19 pandemic and the Russia's war of aggression against Ukraine are particularly motivating for our research. In this paper, we have explored the supply chain robustness under an incomplete information about future shocks. Specifically, the relationship between the firm-level mitigation, and the chain-level preparedness and robustness.

To provide a background of our economy interconnectedness via upstream input sourcing and downstream output markets with authoritarian regimes, we have computed the Foreign Input Reliance and the Foreign Output Reliance with respect to China. To measure the international fragmentation of production processes, we have relied on recent approaches from the theoretical macro literature, which has developed structural interpretations of the world input-output tables. Our descriptive results confirmed that China is an important supplier of inputs to manufacturing industries in all seven largest Alliance's economies. Moreover, the upstream dependency on Chinese inputs has increased during the last two decades. In contrast, the China's reliance on inputs from the Alliance is considerably lower and it has decreased during the last two decades. This dependency on authoritarian regimes' inputs and outputs has to been seen in the context of GSC ruptures. The computed Global Supply Chain Pressure index is on all-time high since 2020. Our first conclusion is that many supply chains across the globe have become not only highly interdependent and complex but also fragile and vulnerable.

A major source of structural problems reducing the resilience of GSCs is a wedge between individual firms' optimal input sourcing paths and a chain-level optimal upstream sourcing diversification. In line with the observation that the society cares relatively more about uncertainty/resilience than private sector firms – which prefer more risk for any given level of reward – we have demonstrated in counterfactual simulations that a chain-level coordination of upstream input sources can qualitatively improve the robustness outcomes of a supply chain. Counterfactual simulations suggest that in the pursuit of efficiency at the firm-level, a supply chain can become vulnerable to aggregate shocks. On the other hand, a supply chain with greater robustness to shocks has to sacrifice efficiency during normal times (in absence of shocks). Our results indicate that an input source diversification extracts a cost in foregone efficiency. While the central planner's first-best solution features a robust supplier sourcing, individual firms do not internalise the benefits of the chain-level robustness and under-diversify upstream suppliers in market economies.

Our findings have direct recommendations to decision makers and to policy choices. Since GSC disruptions in critical sectors may have catastrophic impacts on the social welfare, and the probability of shocks such as Covid-19 and the full-fledged Russia's war of aggression against Ukraine may not be known even approximately, robust decision rules seem to be the appropriate tools for a policy making in critical and strategic sectors such as energy supplies, food and water, communication, and defence. Counterfactual results suggest that a robust supply chain is one in which the survival probability is maximised – as studied in the ambiguity-averse central planner's scenario. A robust decision making ensures that in the presence of an aggregate shock – independently of its realisation – part of input suppliers will survive and the final's goods supplies will continue even under the most demanding circumstances.

Among limitations not addressed in our paper, the presented analysis has not incorporated learning and belief updating mechanisms, which may change the optimal resource allocation when more precise information becomes available to agents. Theoretical ambiguity models with learning and belief updating that we have reviewed in the paper show that the optimal agents' choices depend not only on the content of information but also on the information precision. As the available information precision changes continuously, the agents' belief set is updated and the optimal strategies e.g. to invest in future shock mitigation adjust dynamically even holding the shock probability distribution constant. Implementing empirically and studying these mechanisms in a general equilibrium setup with production are promising avenues for a future research.